\documentclass[%
      aps,
      prd,
      floatfix,
      preprintnumbers,
      preprint,
      showpacs,
      nofootinbib,
      tightenlines,
      amssymb,
      amsmath
]{revtex4-1}

\usepackage{url}
\usepackage{ifpdf}
\ifpdf
	\usepackage{cmap} 
	\usepackage[pdftex]{color,graphicx}
	\usepackage{epstopdf} 
\else
	\usepackage{color,graphicx}
\fi
\usepackage{bm}
\usepackage[dvipsnames]{xcolor}
\usepackage{hyperref}
\hypersetup{%
colorlinks=true,%
citecolor=Blue,%
filecolor=Black,%
linkcolor=Blue,%
urlcolor=Violet
}
\ifpdf
\usepackage[activate={true,nocompatibility},final,tracking=true,kerning=true,spacing=true,factor=1100,stretch=10,shrink=10]{microtype}
\fi

\newcommand{\eq}[1]{Eq.(\ref{#1})}

\begin{document}
\title{
A Precise Determination of (Anti)neutrino Fluxes with 
(Anti)neutrino-Hydrogen Interactions
}

\author{H.~Duyang, B.~Guo, S.R.~Mishra and R.~Petti}
\affiliation{Department of Physics and Astronomy, University of South Carolina, Columbia, South Carolina 29208, USA}


\begin{abstract}

We present a novel method to accurately determine the flux of neutrinos and antineutrinos, 
one of the dominant systematic uncertainty affecting current and   
future long-baseline neutrino experiments, as well as precision neutrino scattering experiment. 
Using exclusive topologies in $\nu(\bar \nu)$-hydrogen interactions, 
$\nu_\mu p \to \mu^- p \pi^+$, $\bar \nu_\mu p \to \mu^+ p \pi^-$, 
and $\bar \nu_\mu p \to \mu^+ n$ with small hadronic energy, 
we achieve an overall accuracy on the relative fluxes 
better than 1\% in the energy range covering most of the available flux. 
Since we cannot rely on simulations nor model corrections at this level of precision, 
we present techniques to constrain all relevant systematic uncertainties 
using data themselves. 
The method can be implemented using the approach we recently proposed to collect 
high statistics samples of $\nu(\bar \nu)$-hydrogen interactions 
in a low-density and high-resolution detector, which could serve as part of the near detector 
complex in a long-baseline neutrino experiment, as well as a 
dedicated beam monitoring detector.

\end{abstract}


\maketitle

\section{Introduction}
\label{sec:intro}

The unprecedented intensity available at modern wide-band (anti)neutrino facilities 
allows the use of high resolution detectors with a relatively small fiducial mass of a few tons 
to achieve an accurate reconstruction of (anti)neutrino interactions, alleviating one of the 
primary limitations of past experiments. 
However, unlike charged lepton scattering experiments, the (anti)neutrino probe 
has to face the intrinsic limitation that the energy of the incoming (anti)neutrino 
is unknown on an event-by-event basis. Cross-sections and fluxes are thus folded into the 
observed event distributions and have to be determined  from the same data. 
Since (anti)neutrino experiments need to use nuclear targets to collect a  
sizable statistics, nuclear effects introduce a substantial smearing on the measured distributions, 
resulting in additional systematic uncertainties. For these reasons all (anti)neutrino scattering 
experiments have been limited by a poor knowledge of the incident flux. 
An accurate knowledge of the (anti)neutrino flux is a necessary condition to exploit the 
unique features of the (anti)neutrino probe for precision measurements of fundamental 
interactions. The flux uncertainties are also the dominant systematic uncertainties in 
current and future long-baseline neutrino oscillation experiments. 

In Ref.~\cite{Duyang:2018lpe} we proposed a novel approach to 
detect $\nu(\bar \nu)$-hydrogen interactions via subtraction between 
dedicated CH$_2$ and graphite (pure C) targets, embedded within a low 
density tracker allowing a control of the target configuration, composition, and mass 
similar to electron scattering experiments. 
We used a kinematic selection -- largely based upon the transverse momentum 
vectors of emergent particles -- 
to precisely identify $\nu(\bar \nu)$-H interactions, 
achieving efficiencies exceeding 90\% and purities of 80-92\%. 
The measurement of the $\nu(\bar \nu)$-C background is entirely data-driven.
This concept -- both simple and safe to implement -- appears to be the only    
realistic opportunity to obtain high statistics samples of 
(anti)neutrino interactions on hydrogen, since safety and practical arguments 
make other techniques unfeasible.

In this paper we propose to use the 
exclusive $\nu_\mu p \to \mu^- p \pi^+$, $\bar \nu_\mu p \to \mu^+ p \pi^-$, 
and $\bar \nu_\mu p \to \mu^+ n$ processes on hydrogen to achieve accurate 
measurements of (anti)neutrino fluxes without the limitations arising from the 
nuclear smearing in conventional targets. 
Furthermore, by restricting the flux measurement to events with small hadronic energy 
we significantly reduce the systematic uncertainties on the energy dependence 
of the cross-sections. 
We perform a detailed analysis using 
realistic assumptions for the detector smearing and physics modeling in order to evaluate the 
relevant uncertainties and the overall precisions achievable by the proposed techniques. 

This paper is organized as follows In Sec.~\ref{sec:analysis} we briefly describe the 
detection technique and the selection of the exclusive event samples for the flux 
determination. In Sec.~\ref{sec:results} we discuss our results and in 
Sec.~\ref{sec:sum} we summarize our findings.

\section{Detection Technique and Event Selection} 
\label{sec:analysis} 

We consider the detection technique proposed in Ref.~\cite{Duyang:2018lpe} to obtain 
$\nu(\bar \nu)$-H interactions from the subtraction of events in dedicated 
CH$_2$ plastic and graphite (pure C) targets. The key detector element is a 
low-density ($\rho \sim 0.16$ g/cm$^3$) 
straw tube tracker, in which thin layers of various target materials (100\% chemical purity) 
are alternated with straw layers so that they represent more than 95\% of the total 
detector mass (5\% being the mass of the straws). As discussed in 
Ref.~\cite{Duyang:2018lpe} this design allows a control of the configuration, chemical composition, size, and mass of the (anti)neutrino targets in a way similar to what is typically done in electron scattering experiments. Our analysis is based upon a fiducial mass of 5 tons of CH$_2$ 
-- corresponding to 714 kg of hydrogen -- and over 500 kg of graphite~\cite{Duyang:2018lpe}. 

We simulate (anti)neutrino interactions on CH$_2$, H, and C targets with 
three different event generators: NuWro~\cite{Juszczak:2005zs}, GiBUU~\cite{Buss:2011mx}, and GENIE~\cite{Andreopoulos:2009rq} to check the sensitivity of our 
analysis to the details of the input modeling. 
We generate inclusive Charged Current (CC) interactions including all 
processes available in the event generators -- quasi-elastic (QE), $\Delta(1232)$ 
and higher resonances (RES), non-resonant processes and deep inelastic scattering (DIS) -- 
with input (anti)neutrino spectra 
similar to the ones expected in the Long-Baseline Neutrino Facility (LBNF) and 
in the DUNE experiment~\cite{Acciarri:2015uup,Acciarri:2016ooe}.
We then use the GEANT4~\cite{Agostinelli:2002hh} program to evaluate 
detector effects and apply to all final state particles a parameterized reconstruction 
smearing consistent with the NOMAD data~\cite{Altegoer:1997gv}. 

We assume the same event selection described in Ref.~\cite{Duyang:2018lpe}.
In particular, for the various flux measurements we focus on two exclusive topologies: 
(a) $\nu_\mu p \to \mu^- p \pi^+$ and $\bar \nu_\mu p \to \mu^+ p \pi^-$, 
mainly from resonance production; 
(b) $\bar \nu_\mu p  \to \mu^+ n$ quasi-elastic interactions. 
The high resolution of the detector we consider allows to identify the   
interactions on hydrogen within the CH$_2$ target by using a kinematic analysis. 
Since the H target is at rest, the Charged Current (CC) events are expected to be 
perfectly balanced in a plane transverse to the beam direction (up to the tiny beam divergence) 
and the muon and hadron vectors are back-to-back in the same plane. Instead, events from 
nuclear targets are affected by both initial and final state nuclear effects, resulting in 
a significant missing transverse momentum and a smearing of the transverse plane kinematics. 
We can exploit these differences using the reconstructed event kinematics. 
The $\mu^- p \pi^+ (\mu^+ p \pi^-)$ samples are selected with 
an efficiency of 90\% and a purity of 92\% (88\%) for $\nu (\bar \nu)$-H, while the 
$\mu^+ n$ QE sample is selected with a purity of 80\%~\cite{Duyang:2018lpe}.  
The distributions of the generic kinematic variables 
$\vec x \equiv (x_1, x_2, ....., x_n)$ in $\nu(\bar \nu)$-H interactions are then obtained as: 
\begin{equation} 
N_H(\vec x) \equiv N_{CH_2} (\vec x) - N_{C} (\vec x) \times \frac{M_{C/CH_2}}{M_C} 
\label{eq:Hevt} 
\end{equation} 
where $N_{CH_2}$ and $N_C$ are the data from the CH$_2$ plastic and graphite (C) 
targets. The interactions from this latter are normalized by the ratio between the total fiducial 
masses of C within the graphite and CH$_2$ targets, $M_{C/CH_2}/M_C$. 
The subtraction in \eq{eq:Hevt} is performed after all the selection cuts including the 
kinematic analysis. In this paper we assume as input the corresponding 
$\nu(\bar \nu)$-H samples obtained after the kinematic analysis 
and the subtraction of the small residual C background using the dedicated graphite target. 
In Sec.~\ref{sec:subsys} we will discuss the additional uncertainties introduced by the 
subtraction procedure on the flux measurements.

\section{Results and Discussion} 
\label{sec:results} 

\subsection{\boldmath Relative $\nu_\mu$ flux} 
\label{sec:numurel} 

Relative fluxes as a function of the (anti)neutrino energy $E_\nu$ have been determined by 
many modern neutrino experiments by using the measured inclusive CC interactions 
with small visible hadronic energy $\nu$~\cite{Mishra90,Bodek:2012uu}. 
This technique (low-$\nu$) is based on the observation that introducing 
a fixed $\nu_0$ cut reduces the available phase space and the corresponding 
energy dependence of the cross-section. 
This latter can be expanded in series of the ratio $\nu_0 / E_\nu$, so that the 
number of observed events with hadronic energy $\nu<\nu_0$ can be written as 
$N(E_\nu, \nu<\nu_0) \; = \; k \Phi(E_\nu) f_c(\nu_0/E_\nu)$, 
where $\Phi$ is the (anti)neutrino flux, $k$ is an arbitrary normalization constant, and 
$f_c$ is a correction factor which can be calculated as a power series in $\nu_0/E_\nu$ 
with coefficients given by combinations of integrals of the structure functions. In practice 
the factor $f_c$ can be evaluated using the MC as the ratio of the cross-section 
with $\nu<\nu_0$ with respect to its asymptotic value at the highest energy of 
interest for the measurement. The correction factor $f_c$ becomes smaller by lowering the 
value of the cut $\nu_0$ and typically gives reliable flux predictions for $E_\nu \gtrsim 2 \nu_0$. 
The use of low energy (anti)neutrino beams for long-baseline oscillation experiments 
requires to use $\nu_0$ cuts in the range 0.25--0.50 GeV~\cite{Bodek:2012uu} and 
the corresponding flux samples are almost entirely composed of quasi-elastic and 
resonant interactions. 

Past and current neutrino experiments have used the low-$\nu$ approach with  
nuclear targets ranging from C to Pb.
The use of such nuclear targets intrinsically limit the accuracy achievable in the 
determination of relative fluxes, due to the 
systematic uncertainties associated to the nuclear smearing including 
Fermi motion and binding, off-shell corrections, meson exchange currents, 
nuclear shadowing~\cite{Kulagin:2004ie,Kulagin:2007ju,Kulagin:2014vsa}, 
neutron production, and final state interactions~\cite{Alvarez-Ruso:2017oui}. 
The nuclear smearing directly affects the hadronic energy 
reconstruction and the acceptance of the cut $\nu<\nu_0$. 

\begin{figure}[t]
\begin{center}
\includegraphics[width=1.00\textwidth]{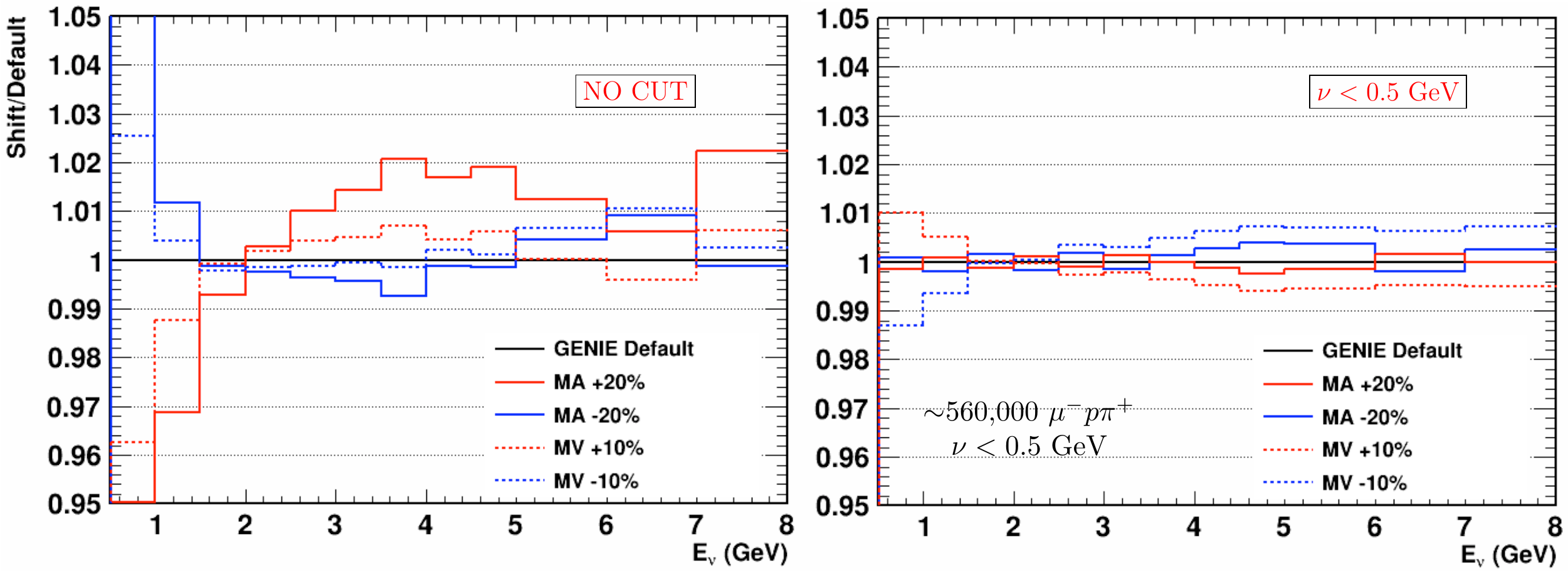}
\end{center}
\caption{Left panel: effect of a variation of the axial and vector form factors on the 
determination of the relative flux from $\nu_\mu p \to \mu^- p \pi^+$ processes on hydrogen. 
Right panel: same as the previous one but with a cut $\nu<0.5$ GeV applied.
}
\label{fig:Hsel-RelFlux3trk}
\end{figure}

\subsubsection{Exclusive $\nu_\mu p \to \mu^- p \pi^+$ on Hydrogen} 
\label{sec:numu3trk} 

The limitations discussed above can be overcome by considering a single exclusive process 
on an elementary target like hydrogen (free proton). The use of a single process rather than 
an inclusive sample offers the advantage of a well defined cross-section, while the availability 
of a hydrogen target eliminates the bottleneck arising from nuclear effects. 
As a result, hadronic uncertainties in the determination of relative fluxes can be 
dramatically reduced. 

The simplest topology available in $\nu$-H interactions is the process 
$\nu_\mu p \to \mu^- p \pi^+$, dominated by resonance production. 
Since all final state particles can be accurately reconstructed in the  
low-density tracker described in Sec.~\ref{sec:analysis}, the unfolding of the detector 
response is controlled by the momentum resolution $\delta p/p\sim 3\%$. 
These features make the $\nu_\mu p \to \mu^- p \pi^+$ topology an excellent tool for the 
determination of the relative $\nu_\mu$ fluxes as a function of $E_\nu$. 

The relevant model uncertainties are the ones affecting the energy dependence of the 
RES cross-section on hydrogen, which is controlled by the proton form factors. 
These uncertainties are substantially smaller than in any nuclear target, 
due to the absence of nuclear effects. In order to estimate their effect on the 
determination of the relative fluxes we vary the axial and vector form factors 
in the event generators and repeat our analysis. The results shown in 
Fig.~\ref{fig:Hsel-RelFlux3trk} (left plot) indicate flux shape uncertainties of the order 
of 2--5\% depending upon the neutrino energy considered. 
We can further reduce such uncertainties by restricting our analysis to events with 
low hadronic energy $\nu$. Given the typical invariant mass of resonant processes, 
cuts down to $\nu<0.5$ GeV are feasible. Figure~\ref{fig:Hsel-RelFlux3trk} (right plot) 
demonstrates that the use of this cut with $\nu_\mu p \to \mu^- p \pi^+$ events on H 
can reduce the hadronic uncertainties on 
the flux determination to the sub-percent level. This effect arises from the flattening 
of the energy dependence of the RES cross-section at $\nu<0.5$ GeV, associated to 
the reduced phase space, which is pushing the residual rise at energies lower 
then the range of interest for the flux measurement. Considering an input flux 
similar to DUNE and the exposures from Ref.~\cite{Duyang:2018lpe}, the 
overall efficiency of the cut $\nu<0.5$ GeV on the reconstructed hadronic energy 
is about 25\% for the $ \mu^- p \pi^+$ topologies on H (Tab.~\ref{tab:effnucut}), 
resulting in a total of 560,000 events expected in the flux sample. 

\begin{figure}[tb]
\begin{center}
\includegraphics[width=1.00\textwidth]{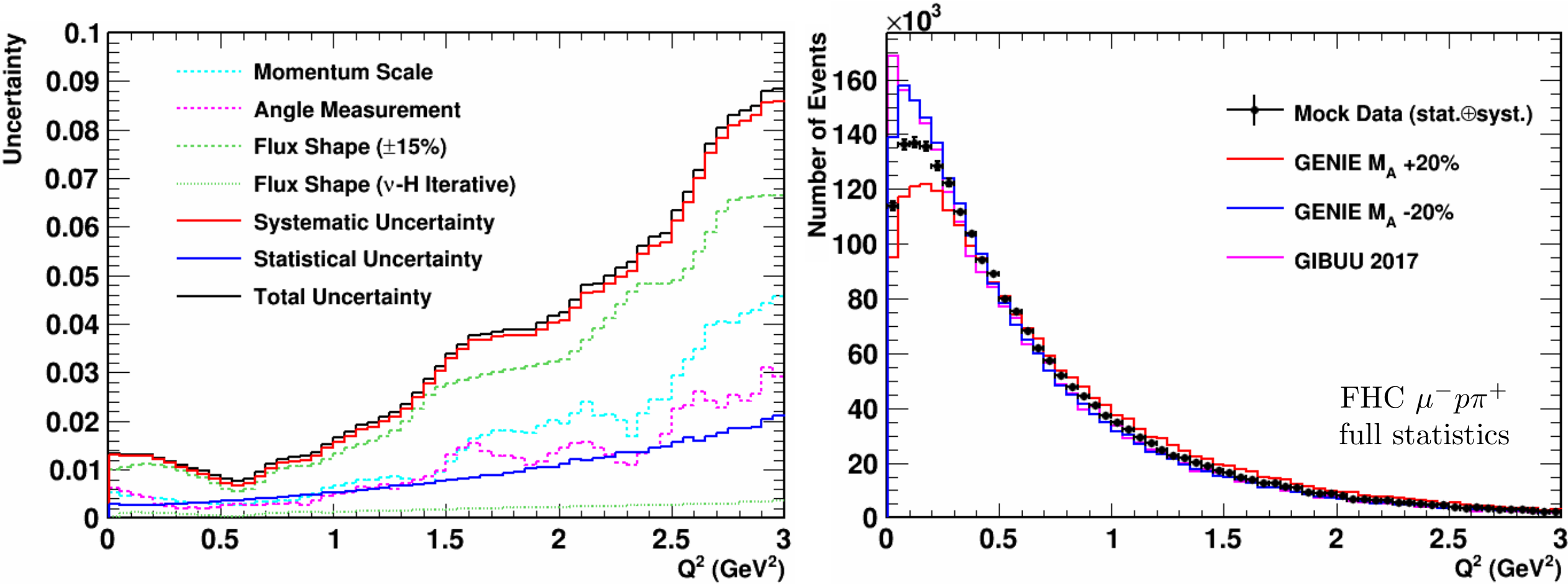}
\end{center}
\caption{Left plot: expected statistical and systematic uncertainties on the 
reconstructed $Q^2$ distribution of $\nu_\mu p \to \mu^- p \pi^+$ events on H. 
Two results for the flux uncertainties are shown: (a) using the initial $\pm 15\%$ uncertainty; 
(b) after an iterative procedure using $\nu_\mu p \to \mu^- p \pi^+$ 
on H with $\nu<0.5$ GeV (Fig.~\ref{fig:Hsel-NumuFlux}). 
Right  plot: reconstructed $Q^2$ distribution of selected 
$\nu_\mu p \to \mu^- p \pi^+$ on H for the complete sample without $\nu$ cut. 
The solid circles (mock-data) correspond to the nominal GENIE cross section
and include both statistical and systematic uncertainties added in quadrature (left plot). 
The sensitivity to a modification of the axial form factor by $M_A \pm 20\%$ is shown 
for illustration purpose, together with the result of the nominal GiBUU simulation. 
All distributions are normalized to the same integral. 
}
\label{fig:Hsel-Q2dist3trk}
\end{figure}
 
As discussed in Sec.~\ref{sec:analysis} , our analysis is based upon inclusive CC 
samples with all relevant processes -- QE, $\Delta(1232)$ and higher resonances, 
non-resonant processes and deep inelastic scattering (DIS) -- for both CH$_2$ and C targets. 
The effects of resonances higher than $\Delta(1232)$ and non-resonant backgrounds 
are reduced by the kinematic selection described in Ref.~\cite{Duyang:2018lpe} 
and are further suppressed by the cut on the hadronic energy $\nu<0.5$ GeV. 
As a result, in the sample used for the 
relative flux determination 96.6\% of the events have $W<1.35$ GeV,  
0.02\% originate from higher mass resonances, and about 3.4\% 
from non-resonant contributions, according to the GENIE simulations. 
Comparable results are obtained 
using the GiBUU and NuWro generators. 
 
We have shown that large variations of the proton form factors result in small 
uncertainties on the relative flux determination from $\nu_\mu p \to \mu^- p \pi^+$ 
interactions on H at $\nu<0.5$ GeV. To this end, in Fig.~\ref{fig:Hsel-RelFlux3trk} we 
consider variations of the vector mass by $\pm$10\% and of the axial mass by $\pm$20\%. 
We emphasize that this estimate is used only for illustration purpose. 
At the level of accuracy (sub-percent) shown in Fig.~\ref{fig:Hsel-RelFlux3trk}  
we cannot rely upon simulations nor model corrections. 
Instead, we will constrain all relevant model uncertainties affecting the 
flux determination using data themselves, as discussed in 
Sec.~\ref{sec:Q2fit}.

\subsubsection{Constraining Model Uncertainties from the $Q^2$ Distribution} 
\label{sec:Q2fit} 

The current theoretical understanding of (anti)neutrino-induced single pion 
production on elementary targets like hydrogen (free proton) is still 
somewhat incomplete~\cite{Alvarez-Ruso:2017oui}. 
Critical aspects are the impact of non-resonant backgrounds and the parametrization 
of the various nucleon form factors involved in the transitions to 
$\Delta (1232)$ and higher resonant states~\cite{Nakamura:2015rta, Hernandez:2016yfb}. 
While substantially smaller than in 
any nuclear target, the corresponding uncertainties suffer from the lack of 
precise measurements besides the limited statistics collected by old bubble 
chamber experiments. Furthermore, many Monte Carlo simulation packages 
rely upon oversimplified model implementations, which are affected by even 
larger uncertainties. 
For all these reasons, the estimates illustrated in Fig.~\ref{fig:Hsel-RelFlux3trk} 
may not cover more general variations of the form factors 
-- which cannot be simply described in terms of axial and vector masses~\cite{Bhattacharya:2011ah} -- 
nor larger unexpected discrepancies with existing models. 

\begin{figure}[tb]
\begin{center}
\includegraphics[width=0.60\textwidth]{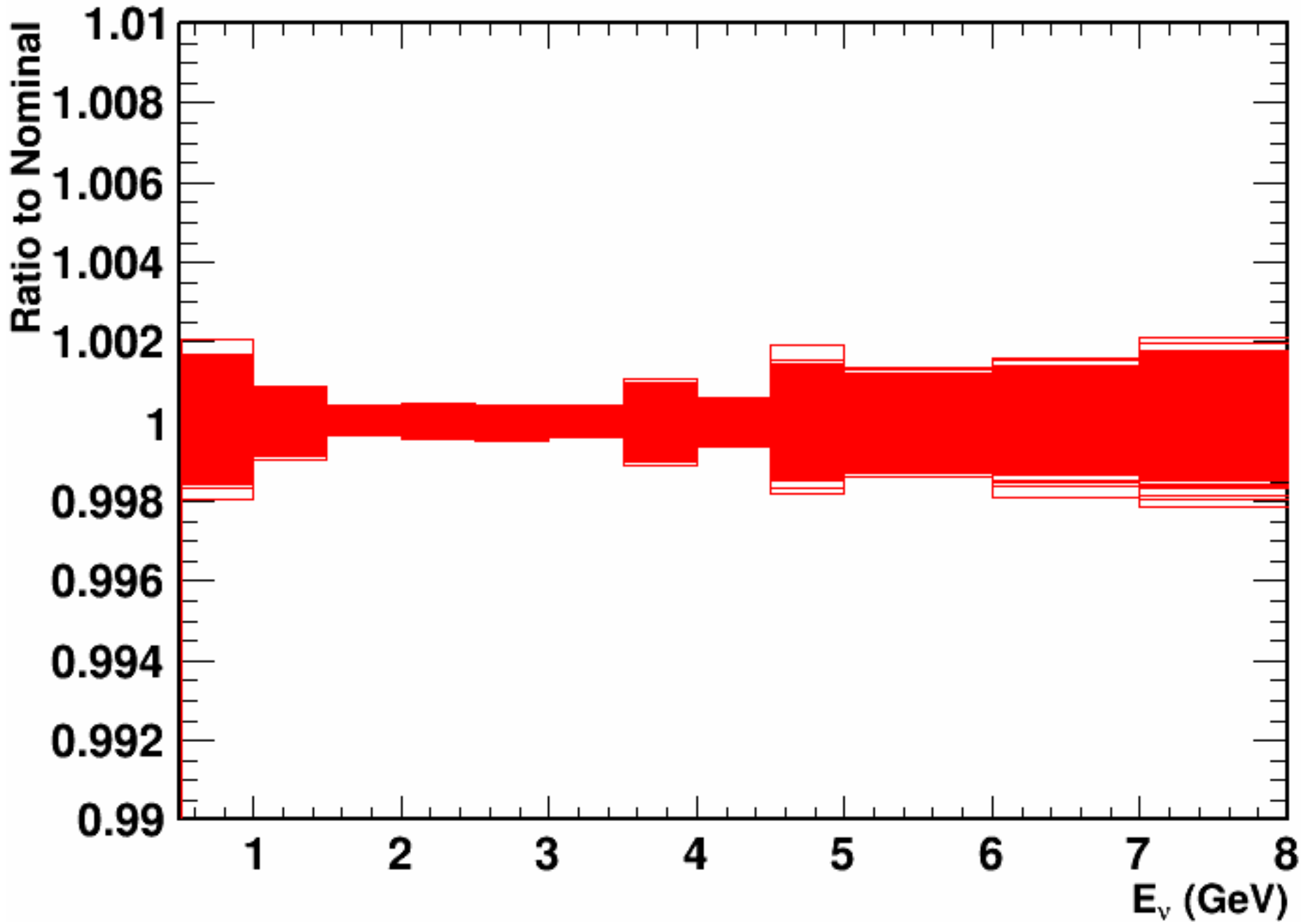}
\end{center}
\caption{Uncertainty on the relative $\nu_\mu$ flux resulting from the 
total statistical and systematic uncertainties (added in quadrature) on the reconstructed 
$Q^2$ distribution for the $\nu_\mu p \to \mu^- p \pi^+$ events on H (Fig.~\ref{fig:Hsel-Q2dist3trk}).  
Initial flux uncertainties ($\pm 15\%$) without the iterative procedure are used. 
The curves represent the result of 1,000 experiments in which the measured $Q^2$ distribution 
is allowed to vary bin-by-bin by $\pm1\sigma$ in a model-independent 
way. See text for details. 
}
\label{fig:Hsel-FluxErrorFromQ2}
\end{figure}

We can address the issues above in a model-independent way by directly analyzing 
the reconstructed $Q^2$ distribution in the complete $\nu_\mu p \to \mu^- p \pi^+$ 
sample on H without the $\nu$ cut. 
Since form factors are expected to be a function of $Q^2$, any modification affecting the 
energy dependence of the cross-section relevant for the relative flux determination 
would manifest as a distortion in the measured $Q^2$ distribution. 
Using the exposures from Ref.~\cite{Duyang:2018lpe}, the total statistics expected is 
about $2.24 \times 10^6$ selected $\mu^- p \pi^+$ events on H. 
This statistics provides a stringent test against arbitrary model variations and a 
good $Q^2$ coverage to directly extract the relevant (effective) proton form factors 
from data themselves. The fraction of overlap events between the flux sample with 
$\nu<0.5$ GeV and the total $\mu^- p \pi^+$ sample is about 
25\% (Tab.~\ref{tab:effnucut}), allowing a robust in-situ 
measurement~\footnote{A similar analysis can be applied to the measured $\nu$ distribution.}. 

In order to evaluate the sensitivity of the reconstructed $Q^2$ distribution to model 
variations, we perform a detailed study of the corresponding systematic uncertainties, 
which are expected to be dominant over statistical uncertainties for the available exposures. 
We consider three main sources of systematic uncertainties: (i) energy dependence of the 
neutrino flux; (ii) momentum scales; (iii) muon angle reconstruction. The first effect is 
particularly important since it can potentially interfere with the possibility to use 
the measured $Q^2$ distribution to constrain systematic uncertainties on the 
relative $\nu_\mu$ flux itself, as determined from events with $\nu<0.5$ GeV. 
For all our studies we use only shape information and normalize 
all $Q^2$ distributions to unit area. The normalization constraint is useful to 
focus on the effect of form factors, reducing the correlation with the 
total cross-section at the expense of some statistical power. 

We assume an initial flux uncertainty of 15\% at all energy values, which is 
significantly larger than the estimates obtained from beam simulations. 
We then simulate 1,000 experiments randomly varying each energy bin by $\pm1\sigma$ 
and take the outer envelope of all the corresponding variations on the $Q^2$ distribution 
as systematic uncertainty. The result is shown in Fig.~\ref{fig:Hsel-Q2dist3trk} (left plot). 
We emphasize that this approach provides an upper limit on the flux uncertainties, 
since these latter can be dramatically reduced by using an iterative 
procedure with the relative flux uncertainties determined in-situ from 
the $\nu_\mu p \to \mu^- p \pi^+$ sample on H with $\nu<0.5$ GeV (Fig.~\ref{fig:Hsel-NumuFlux}). 
For the uncertainties related to the momentum scales and the muon angle 
reconstruction we use the values achieved by the NOMAD experiment from 
$K_0 \to \pi^+\pi^-$ and $\Lambda \to p \pi^-$ decays, as discussed in Sec.~\ref{sec:syst}. 
Figure~\ref{fig:Hsel-NumuFlux} summarizes the statistical and systematic uncertainties 
on the measured $Q^2$ distribution. The rise of systematic uncertainties visible at 
$Q^2>1$ GeV$^2$ is largely the effect of the normalization constraint on a region 
populated with relatively small statistics. 

The sensitivity of the measured $Q^2$ distribution -- including statistical and systematic 
uncertainties added in quadrature -- to model variations is illustrated in 
Fig.~\ref{fig:Hsel-Q2dist3trk} (right plot) for the $\nu_\mu p \to \mu^- p \pi^+$ sample on H. 
The same variations of form factors resulting 
in sub-percent uncertainties on the relative fluxes (Fig.~\ref{fig:Hsel-RelFlux3trk}) 
produce large changes in the shape of the measured $Q^2$ distribution: 
changing the axial mass by $+20\%(-20\%)$ results in a $\chi^2$/dof of 1464/60 (1187/60), 
without using the iterative procedure for the dominant flux uncertainties. 
The difference between the GENIE and GiBUU implementations is   
also distinguishable ($\chi^2$/dof of 2211/60). We perform a model-independent study of the 
constraints obtained from the $Q^2$ distribution on the 
relative $\nu_\mu$ flux determination by randomly varying each $Q^2$ bin 
by $\pm1\sigma$ of the total uncertainty (Fig.~\ref{fig:Hsel-Q2dist3trk}). 
We simulate 1,000 experiments and estimate the systematic 
uncertainty on the relative flux from the outer envelope containing all such 
variations. The results shown in Fig.~\ref{fig:Hsel-FluxErrorFromQ2} are below 
0.2\% at all energies and are smaller than the initial estimate obtained 
by simply changing the axial and vector masses. 

\begin{table}[tb]
\begin{center} 
\begin{tabular}{|l|c|c|c|c|c|c|} \hline
    &     \multicolumn{2}{|c|}{ $\nu_\mu p \to \mu^- p \pi^+$} 
    &     \multicolumn{2}{|c|}{$\bar \nu_\mu p \to \mu^+ p \pi^-$}  &  
    \multicolumn{2}{|c|}{$\bar \nu_\mu p \to \mu^+ n$}  \\ 
$\nu$ cut (GeV)  &  ~~$\nu<0.50$~~ &  ~~$\nu<0.75$~~  &  ~~$\nu<0.50$~~ &  ~~$\nu<0.75$~~  &  ~~$\nu<0.10$~~ &  ~~$\nu<0.25$~~ \\ \hline\hline
Low energy beam   &    25.2\%  &      &   14.9\% &     &  46.8\%  &   76.0\%     \\ 
High energy beam   &      &    45.6\%  &    &  20.7\%   &   39.3\%  &   68.0\% \\ 
\hline
\end{tabular}
\caption{Average efficiencies of the various $\nu$ cuts for the $\mu p \pi$ and $\mu^+ n$ 
QE topologies on hydrogen used in our analysis for the two beam configurations considered.} 
\label{tab:effnucut} 
\end{center} 
\end{table}

\subsection{\boldmath Absolute $\nu_\mu$ flux} 
\label{sec:numuabs} 

The $\nu e^- \to \nu e^-$ elastic scattering offers a purely leptonic process with well 
understood cross-section to be used for the determination of the absolute 
$\nu_\mu$ flux~\footnote{The $\nu e^- \to \nu e^-$ process can also 
provide some information on the relative flux, free from nuclear effects. 
However, the limited statistics and the additional smearing associated to the 
outgoing neutrino and the beam divergence result in much larger uncertainties 
compared to the ones achievable with $\nu(\bar \nu)$-H interactions (Sec.~\ref{sec:syst}).}. 
The experimental signature is defined by a single forward electron in the final 
state~\cite{Park:2015eqa}. 
This process can be efficiently selected in the detector considered in Sec.~\ref{sec:analysis} 
thanks to the excellent electron identification capability and angular and momentum resolutions.
By requiring small values of $E_e \theta_e^2<0.0012$ GeV rad$^2$ for the electron we 
obtain an efficiency of about 84\% with a total background of 5\%, composed of 
$\nu_e$ QE interactions without reconstructed proton (3\%) and NC $\pi^0$ interactions (2\%). 
With the exposures considered in Sec.~\ref{sec:stat} we expect more than 4,000 selected 
signal events in the standard low energy beam and about 10,000 signal events in the 
high energy beam option. Systematic uncertainties on the selected $\nu e^- \to \nu e^-$ 
sample are expected to be about 1\% in the low-density detector considered, resulting in 
a total uncertainty on the absolute $\nu_\mu$ flux of 1.9\% (1.4\%) with the low (high) 
energy beam. 

An independent measurement of the absolute $\nu_\mu$ flux could be obtained 
from $\nu_\mu p \to \mu^- p \pi^+$ interactions on H by restricting the analysis to 
low momentum transfer $\mid \vec{q} \mid$ and low energy transfer $\nu$. 
For values close enough to the threshold kinematics: 
\begin{equation} 
\label{eq:thr1pi} 
(m_N + \nu )^2 - \mid \vec{q} \mid^2 > (m_N + m_\pi )^2
\end{equation} 
where $m_N$ is the nucleon mass and $m_\pi$ the pion mass, 
the $\nu_\mu p \to \mu^- p \pi^+$ cross-section can be calculated using the 
covariant chiral perturbation theory approach of Ref.~\cite{Yao:2018pzc}. 
With the exposures considered in Sec.~\ref{sec:stat} a cut $\mid \vec{q} \mid < 350$ MeV/c 
seems feasible, still retaining more than 21,000 events with the low energy beam option. 
The possibility to use this sample to reduce the uncertainties in the chiral perturbation 
calculations and to obtain an alternative absolute flux determination has to be explored~\cite{Luis19}. 

\begin{figure}[tb]
\begin{center}
\includegraphics[width=1.00\textwidth]{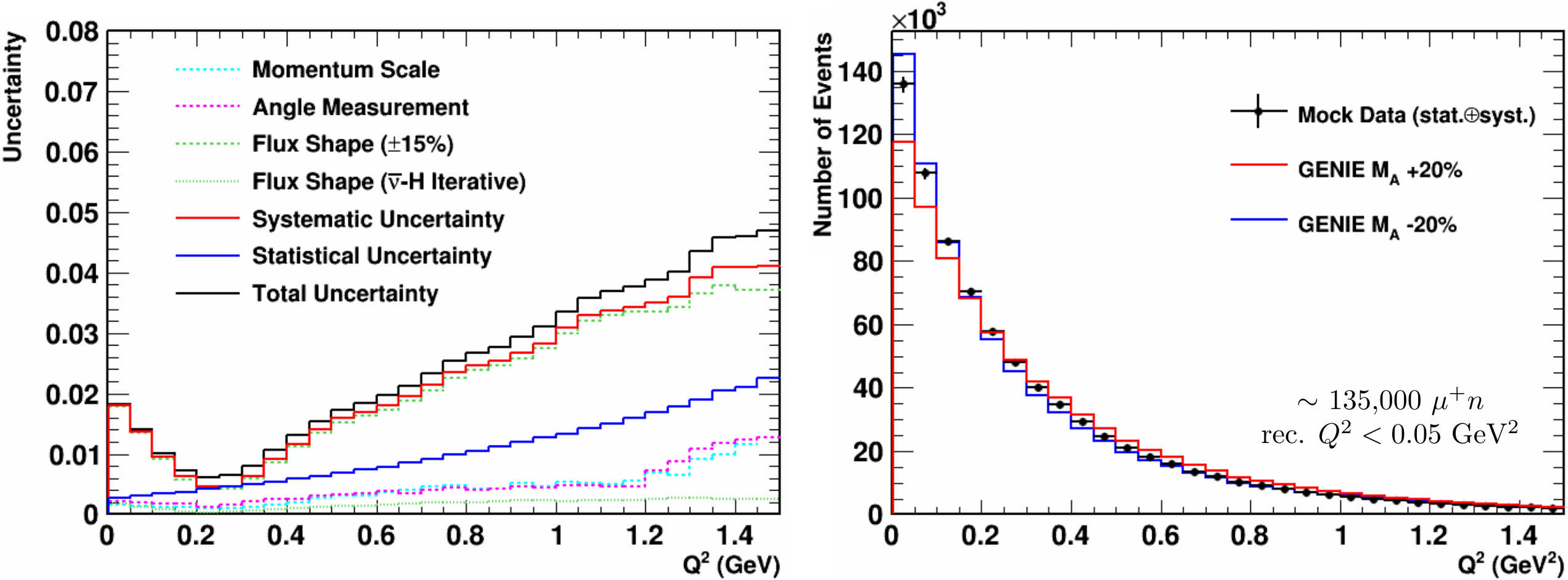}
\end{center}
\caption{Left plot: expected statistical and systematic uncertainties on the 
reconstructed $Q^2$ distribution of $\bar \nu_\mu p \to \mu^+ n$ events on H. 
Two results for the flux uncertainties are shown: (a) using the initial $\pm 15\%$ uncertainty; 
(b) after an iterative procedure using $\bar \nu_\mu p \to \mu^+ n$ 
on H with $\nu<0.25$ GeV (Fig.~\ref{fig:Hsel-NumubarFlux}). 
Right  plot: reconstructed $Q^2$ distribution of selected 
$\bar \nu_\mu p \to \mu^+ n$ on H for the complete sample without $\nu$ cut. 
The solid circles (mock-data) correspond to the nominal GENIE cross section
and include both statistical and systematic uncertainties added in quadrature (left plot). 
The sensitivity to a modification of the axial form factor by $M_A \pm 20\%$ is shown 
for illustration purpose. All distributions are normalized to the same integral. 
}
\label{fig:Hsel-Q2distQE}
\end{figure}

\subsection{\boldmath Relative $\bar \nu_\mu$ flux} 
\label{sec:anumurel} 

The process $\bar \nu_\mu p \to \mu^+ p \pi^-$ on hydrogen has the same 
experimental signature and features as the corresponding $\nu_\mu p$ discussed 
in Sec~\ref{sec:numurel}. We can therefore perform a similar analysis and use 
the sample with $\nu<0.5$ GeV to determine the relative $\bar \nu_\mu$ flux as 
a function of $E_{\bar \nu}$. Considering the exposures from Ref.~\cite{Duyang:2018lpe}, the 
overall efficiency of the cut $\nu<0.5$ GeV on the reconstructed hadronic energy 
is about 15\% for the $ \mu^+ p \pi^-$ topologies on H (Tab.~\ref{tab:effnucut}). 
The efficiency of the $\nu$ cut is lower for the antineutrino samples 
due to the larger contribution from higher 
resonances and non-resonant events to the inclusive $\mu^+ p \pi^-$ topologies. 
Model systematics on the flux determination are similar to the ones discussed in 
Sec~\ref{sec:numurel}.

In addition to the $\bar \nu_\mu p \to \mu^+ p \pi^-$ topologies, we also have the exclusive 
$\bar \nu_\mu p \to \mu^+ n$ QE process in $\bar \nu_\mu$ CC interactions on hydrogen. 
These QE events can be efficiently reconstructed~\cite{Duyang:2018lpe} in the 
detector described in Sec.~\ref{sec:analysis} and can also be used to determine 
the relative $\bar \nu_\mu$ flux in a way similar to the $ \mu^+ p \pi^-$ events. 
The QE sample allows a lower cut on the reconstructed hadronic  energy 
down to $\nu<0.25$ GeV, which has an overall efficiency of 76\% with the 
beam spectrum of Ref.~\cite{Duyang:2018lpe} (Tab.~\ref{tab:effnucut}). 
With the event selection and exposures of Ref.~\cite{Duyang:2018lpe} we 
expect a total of about 812,000 reconstructed QE events on H, 
out which 617,000 have $\nu<0.25$ GeV. 
The model uncertainties on the relative flux can be directly constrained 
by extracting the relevant (effective) form factors from the measured $Q^2$ distribution
(Fig.~\ref{fig:Hsel-Q2distQE}). 
The overlap with the flux sample can be 
reduced below 50\% with a lower cut $\nu<0.1$ GeV (Tab.~\ref{tab:effnucut}). 
The large statistics of the complete reconstructed QE sample without 
the $\nu$ cut provides a good sensitivity to constrain arbitrary model variations, 
following the same approach discussed in Sec.~\ref{sec:Q2fit}. 
Figure ~\ref{fig:Hsel-Q2distQE} (right plot) illustrates this sensitivity: 
changing the axial mass by $+20\%(-20\%)$ results in a $\chi^2$/dof of 828/30 (291/30), 
without using the iterative procedure for the dominant flux uncertainties.

\subsection{\boldmath Absolute $\bar \nu_\mu$ flux} 
\label{sec:anumuabs} 

The availability of large samples of $\bar \nu_\mu p \to \mu^+ n$ QE events on 
hydrogen also allows a determination of the absolute $\bar \nu_\mu$ flux, 
in addition to the relative one as a function of $E_{\bar \nu}$ discussed in Sec~\ref{sec:anumurel}. 
The cross-section for the $\bar \nu_\mu$ QE process on hydrogen 
in the limit of $Q^2 \to 0$ can be written as: 
\begin{equation} 
\label{eq:QE} 
\frac{d\sigma}{dQ^2} \mid_{Q^2=0} \;\; = \;\; \frac{G_F^2 \cos^2\theta_c}{2\pi} \left[ F_V^2(0) + F_A^2 (0) \right]
\end{equation} 
where $F_V$ and $F_A$ are the vector and axial form factors, $\theta_c$ is the Cabibbo angle, 
$G_F$ the Fermi constant, and we have neglected terms in $(m_\mu / M)^2$. The cross-section 
in \eq{eq:QE} 
at $Q^2=0$ is determined by the neutron $\beta$ decay to a precision $\ll 1\%$. 
Experimentally, we can select low $Q^2$ QE events and determine the asymptotic value by 
fitting the measured $Q^2$ distributions (Fig.~\ref{fig:Hsel-Q2distQE}). 
Considering the exposures from Ref.~\cite{Duyang:2018lpe}, 
we expect about 135,000  reconstructed $\bar \nu_\mu p \to \mu^+ n$ 
QE events with $Q^2<0.05$ GeV$^2$ (corresponding to $\nu< 27$ MeV). 
We note that in a detector like the one discussed in Sec.~\ref{sec:analysis} neutrons can be 
detected down to a much lower threshold than protons, thus enhancing the 
reconstruction efficiency of $\bar \nu_\mu$ QE on H at very small $Q^2$ values. 
The measurement of the absolute $\bar \nu_\mu$ flux using QE interactions on H 
requires a calibration of the absolute neutron detection efficiency, 
which can be performed using dedicated test-beam exposures of the relevant detector elements.

\subsection{Flux uncertainties} 
\label{sec:errors} 

\subsubsection{Exposures and Statistical Uncertainties} 
\label{sec:stat} 

In order to study the statistical and systematic uncertainties on the 
$\nu_\mu$ and $\bar \nu_\mu$ fluxes achievable with the method we propose, we consider 
the realistic case study of the fluxes and exposures from Ref.~\cite{Duyang:2018lpe}. 
The same beam and detector assumptions were the basis of a proposal to 
enhance the sensitivity to long-baseline oscillations in LBNF/DUNE and to define an 
extensive program of precision tests of fundamental interactions~\cite{Petti2018,ESGprop}.
As an illustration of the flexibility of the method we consider two different beam spectra with 
the exposures of Ref.~\cite{ESGprop}: 
(a) a low energy beam similar to the default one optimized for the search for CP violation in 
DUNE~\cite{Acciarri:2015uup,Acciarri:2016ooe}; (b) a high energy beam option 
optimized for the $\nu_\tau$ appearance from long-baseline oscillations. 

The statistical uncertainties considered are the ones related to the selection of the 
exclusive $\nu_\mu p \to \mu^- p \pi^+$ and $\bar \nu_\mu p \to \mu^+ n$ QE topologies 
on H described in Sec.~\ref{sec:results}, for the assumed exposures.

\begin{table}[tb]
\begin{center} 
\begin{tabular}{|l|c|c|c|c|} \hline
    &     \multicolumn{2}{|c|}{CP optimized}
    &     \multicolumn{2}{|c|}{$\nu_\tau$ optimized}   \\ 
  &  ~~FHC~~ &  ~~RHC~~  &  ~~FHC~~ &  ~~RHC~~  \\ \hline\hline
~~~$K_0 \to \pi^+\pi^-$~~~   &    ~~~264,000~~~  &   ~~~132,000~~~   &   ~~~1,981,000~~~ &    ~~~665,000~~~    \\ 
~~~$\Lambda \to p \pi^-$   &   ~~~293,000~~~   &   ~~~104,000~~~   &  ~~~1,998,000~~~  &   ~~~503,000~~~  \\ 
\hline
\end{tabular}
\caption{Expected numbers of $K_0$ and $\Lambda$ decays in charged and neutral current interactions with the exposures of Sec~\ref{sec:stat} for the various beam configurations considered.} 
\label{tab:V0s} 
\end{center} 
\end{table} 

\subsubsection{Systematic Uncertainties} 
\label{sec:syst}

We study the effect of three different 
sources of systematic uncertainties on the fluxes determined from $\nu(\bar \nu)$-H 
interactions: (i) muon energy scale; (ii) hadronic energy reconstruction and $\nu$ cut; 
(iii) modeling of form factors and cross-sections. 

Since the flux samples (Sec.~\ref{sec:results}) include events with small hadronic energy  
$\nu$, the dominant contribution to the visible energy comes from the muon. 
The accuracy in determining the muon energy scale is therefore crucial for all the 
flux measurements, requiring a low density tracking detector, as well as a precise calibration 
of the measured momenta for the charged particles. The density of the detector described 
in Sec.~\ref{sec:analysis}, $\rho \sim 0.16$ g/cm$^3$, and its track sampling are well suited 
for these measurements. Following the technique used by the NOMAD 
experiment~\cite{Altegoer:1997gv} 
-- based upon a similar detector concept -- we can calibrate the momentum scale of 
charged particles with the mass peak of the large samples of reconstructed 
$K_0 \to \pi^+\pi^-$ decays~\cite{Wu:2007ab} (Tab.~\ref{tab:V0s}). 
In our study we assume the same muon energy scale uncertainty of 0.2\% achieved 
by the NOMAD experiment~\cite{Wu:2007ab}. We note that the detector we consider 
would provide 25 time higher granularity than NOMAD and about 40 times higher $K_0$ 
statistics, as shown in Tab.~\ref{tab:V0s}. 

The proton reconstruction efficiency and the corresponding energy scale can be accurately 
calibrated with the large samples of $\Lambda \to p \pi^-$ decays available (Tab.~\ref{tab:V0s}). 
Identified $\Lambda$ decays provide a good constrain on systematic uncertainties 
related to the hadronic energy and vertex reconstruction, since the hadron final state 
particles are the same as in the $\nu_\mu p \to \mu^- p \pi^+$ process on hydrogen used 
for the flux determination. Furthermore, both $\Lambda$ and $K_0$ decays can be 
used to constrain the systematic uncertainty on the muon angle reconstruction, which 
is relevant for the analysis of the $Q^2$ distribution discussed in Sec.~\ref{sec:Q2fit}. 

To estimate the effects of the hadronic energy reconstruction on the flux measurements 
we consider a realistic detector smearing and event selection from 
Ref.~\cite{Duyang:2018lpe} (Sec.~\ref{sec:analysis}). 
The acceptance for individual final state particles ($p,n,\pi,\mu$) takes into account the 
detector geometry, the event topology, and the material traversed by the 
particles~\footnote{Protons originated from the process $\nu_\mu p \to \mu^- p \pi^+$ on H 
have a relatively long range in the considered detector ($\rho \sim 0.16$ g/cm$^3$): 
about 99.8\% (87.3\%) of them has $p>100 (200)$ MeV/c.}. 
It is then folded into the analysis and the 
reconstruction smearing on the hadronic energy is evaluated as a function of $\nu$. 
In addition to the detector response and event selection, 
we vary the $\nu$ cut applied to define the flux samples according to the 
expected resolution around the cut values. 
We also study the effect of different $\nu$ cuts in the range 0.25--0.75 GeV
to optimize the sensitivity of the analysis for different beam spectra and 
$\nu(\bar \nu)$-H topologies. 

Model uncertainties are estimated by varying the vector form factor by $\pm$10\% and the 
axial form factor by $\pm$20\%, as described 
in Sec~\ref{sec:numu3trk}. These variations are relatively large and 
provide an upper limit on the corresponding expected 
uncertainties, since we deliberately ignore the in-situ constraints on the form factors 
obtained from the measured $Q^2$ distributions. We emphasize that the realistic 
uncertainties on the flux obtained from a model-independent analysis of the 
$Q^2$ distribution are significantly smaller (Fig.~\ref{fig:Hsel-FluxErrorFromQ2}), 
as discussed in Sec.~\ref{sec:Q2fit}.

For each variation of the relevant parameters within their systematic uncertainties we 
repeat our analysis and evaluate the difference in the extracted fluxes as a function 
of $E_\nu$. To this end, we consider both positive and negative variations and 
use the largest of the two as the corresponding uncertainty. 
Since the determination of relative fluxes is 
defined up to an arbitrary constant, we normalize these measurements 
to unit area by dividing them by the integral of the measured distributions. 
The absolute flux normalization is provided independently by the $\nu e^- \to \nu e^-$ 
elastic scattering for $\nu_\mu$ (Sec.\ref{sec:numuabs}) and by the 
$\bar \nu_\mu p \to \mu^+ n$ QE on H 
at $Q^2=0$ for $\bar \nu_\mu$ (Sec.~\ref{sec:anumuabs}), respectively. 

\begin{figure}[t]
\begin{center}
\includegraphics[width=1.00\textwidth]{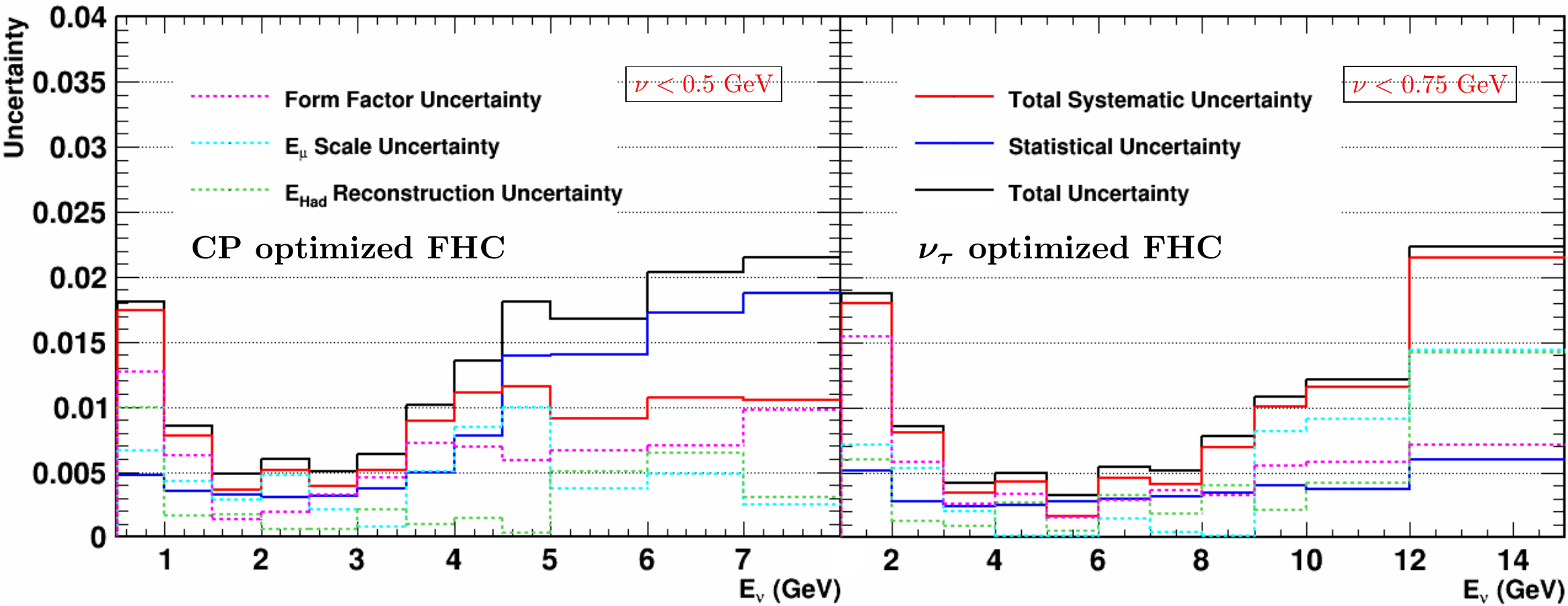}
\end{center}
\caption{Summary of the expected statistical and systematic uncertainties in 
the $\nu_\mu$ relative flux determination using $\nu_\mu p \to \mu^- p \pi^+$ exclusive 
processes on hydrogen. Two different input spectra 
similar to the ones planned in the LBNF are considered: 
(a) a low-energy beam optimized to search for CP violation (left plot) 
with a cut $\nu<0.5$ GeV; 
(b) a high-energy beam optimized to detect the $\nu_\tau$ appearance (right plot)
with a cut $\nu<0.75$ GeV. 
See text for details. 
}
\label{fig:Hsel-NumuFlux}
\end{figure}

The statistical and systematic uncertainties expected on the relative $\nu_\mu$ flux 
determined from $\nu_\mu p \to \mu^- p \pi^+$ interactions on H are shown in 
Fig.~\ref{fig:Hsel-NumuFlux} for both the low energy and high energy beam 
options considered. In the former case we use a cut $\nu<0.5$ GeV (Sec.~\ref{sec:results}), 
while in the latter a higher cut $\nu < 0.75$ GeV turns out to be more appropriate. 
In the energy ranges where we expect the bulk of the fluxes the total uncertainties 
-- including both the statistical and systematic ones added in quadrature -- are well below 1\%. 
The dominant systematic uncertainty is related to the muon energy scale, as 
hadronic and model uncertainties are small for interactions on hydrogen at small 
values of the energy transfer $\nu$. 
The statistical uncertainty is dominating the results 
in the tails of the available spectra. 
Uncertainties with the high energy beam option are smaller than with the 
low energy beam due to the higher statistics and to the broader energy spectrum. 
The level of accuracy on the flux determination 
demonstrated in Fig.~\ref{fig:Hsel-NumuFlux} cannot be achieved by  
other known techniques using nuclear targets. As illustrated by the comparison 
between two different spectra, the proposed method can be easily adapted to 
a wide range of beam configurations, provided the exposures are large enough 
to offer the required  statistics $\mathcal{O}(10^6)$ for the various exclusive samples considered. 

\begin{figure}[t]
\begin{center}
\includegraphics[width=1.00\textwidth]{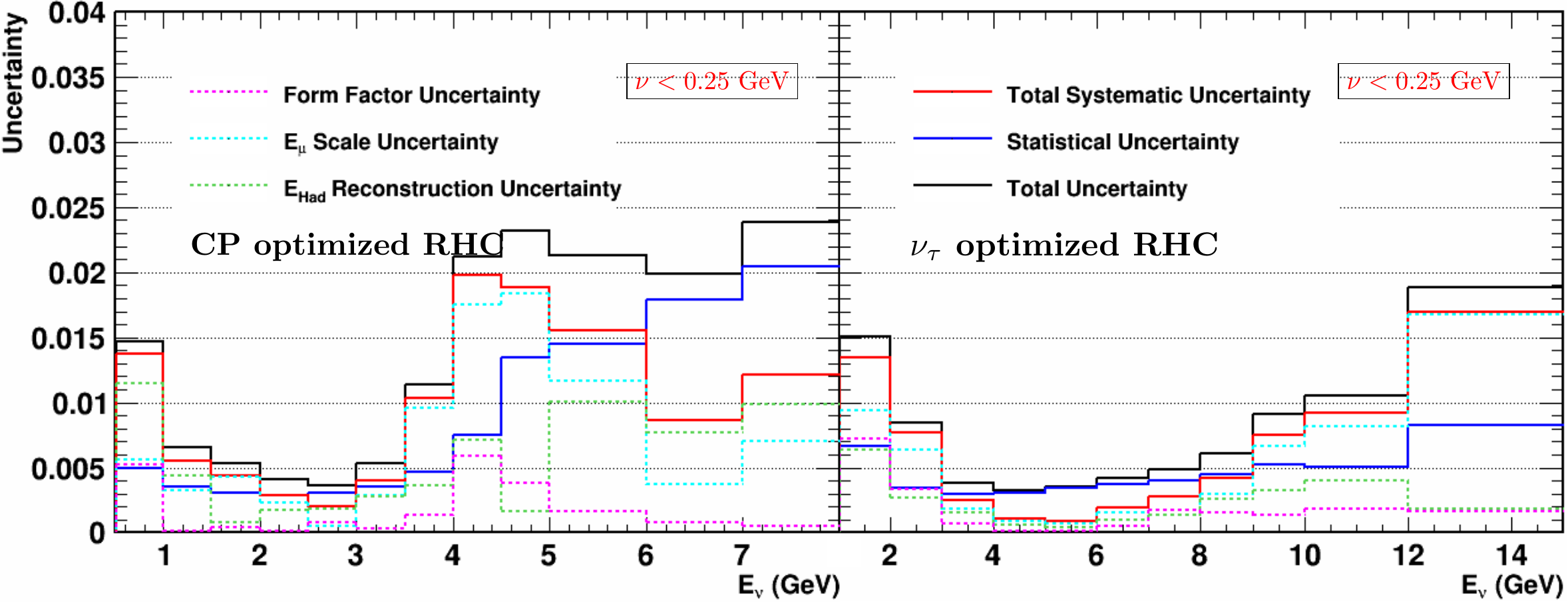}
\end{center}
\caption{Summary of the expected statistical and systematic uncertainties in 
the $\bar \nu_\mu$ relative flux determination using $\bar \nu_\mu p \to \mu^+ n$ 
QE exclusive processes on hydrogen. Two different input spectra 
similar to the ones planned at the LBNF are considered: 
(a) a low-energy beam optimized to search for CP violation (left plot) 
with a cut $\nu<0.25$ GeV; 
(b) a high-energy beam optimized to detect the $\nu_\tau$ appearance (right plot) 
with a cut $\nu<0.25$ GeV. 
See text for details. 
}
\label{fig:Hsel-NumubarFlux}
\end{figure}

Figure~\ref{fig:Hsel-NumubarFlux} shows the statistical and systematic uncertainties 
on the relative $\bar \nu_\mu$ flux determined from $\bar \nu_\mu p \to \mu^+ n$ QE 
interactions on H with $\nu<0.25$ GeV. Given the larger efficiencies (Tab.~\ref{tab:effnucut}), 
we can also apply a cut $\nu<0.1$ GeV for both the low and the high energy 
beam options. Similar considerations as for the relative $\nu_\mu$ fluxes can be made. 

\subsubsection{Effect of the C Background Subtraction}  
\label{sec:subsys} 

The kinematic analysis described in Sec.~\ref{sec:analysis} allows the identification 
of the $\nu_\mu p \to \mu^- p \pi^+$, $\bar \nu_\mu p \to \mu^+ p \pi^-$, and 
$\bar \nu_\mu p \to \mu^+ n$ topologies  
within the CH$_2$ target with little residual backgrounds $\sim$8-20\% from 
interactions on the carbon nucleus~\footnote{We obtain similar efficiencies and 
purities using three independent event generators: 
NuWro, GiBUU, and GENIE~\cite{Duyang:2018lpe}.}. 
The purity of the H samples can be further increased 
by tightening the multivariate selection~\cite{Duyang:2018lpe}. 
A necessary condition to reduce systematic 
uncertainties on the subtraction of the small C background is to use a model-independent 
approach based entirely upon the data obtained from the dedicated graphite (pure C) target. 
The detector technology discussed in Sec.~\ref{sec:analysis} is essential, 
since the CH$_2$ and C targets are configured as thin layers, 
ensuring that the corresponding acceptance 
corrections are small and, most importantly, similar for both targets. 
We verified this latter condition with detailed detector simulations using the 
GEANT4 program~\cite{Agostinelli:2002hh}. We emphasize that the data from 
the graphite target automatically include all types of interactions, 
as well as reconstruction effects, relevant for our analysis. 
The impact of possible model dependencies through the acceptance corrections 
is therefore negligible, since they would appear as third order effects on 
the data-driven subtraction of small backgrounds. 

\begin{figure}[tb]
\begin{center}
\includegraphics[width=0.55\textwidth]{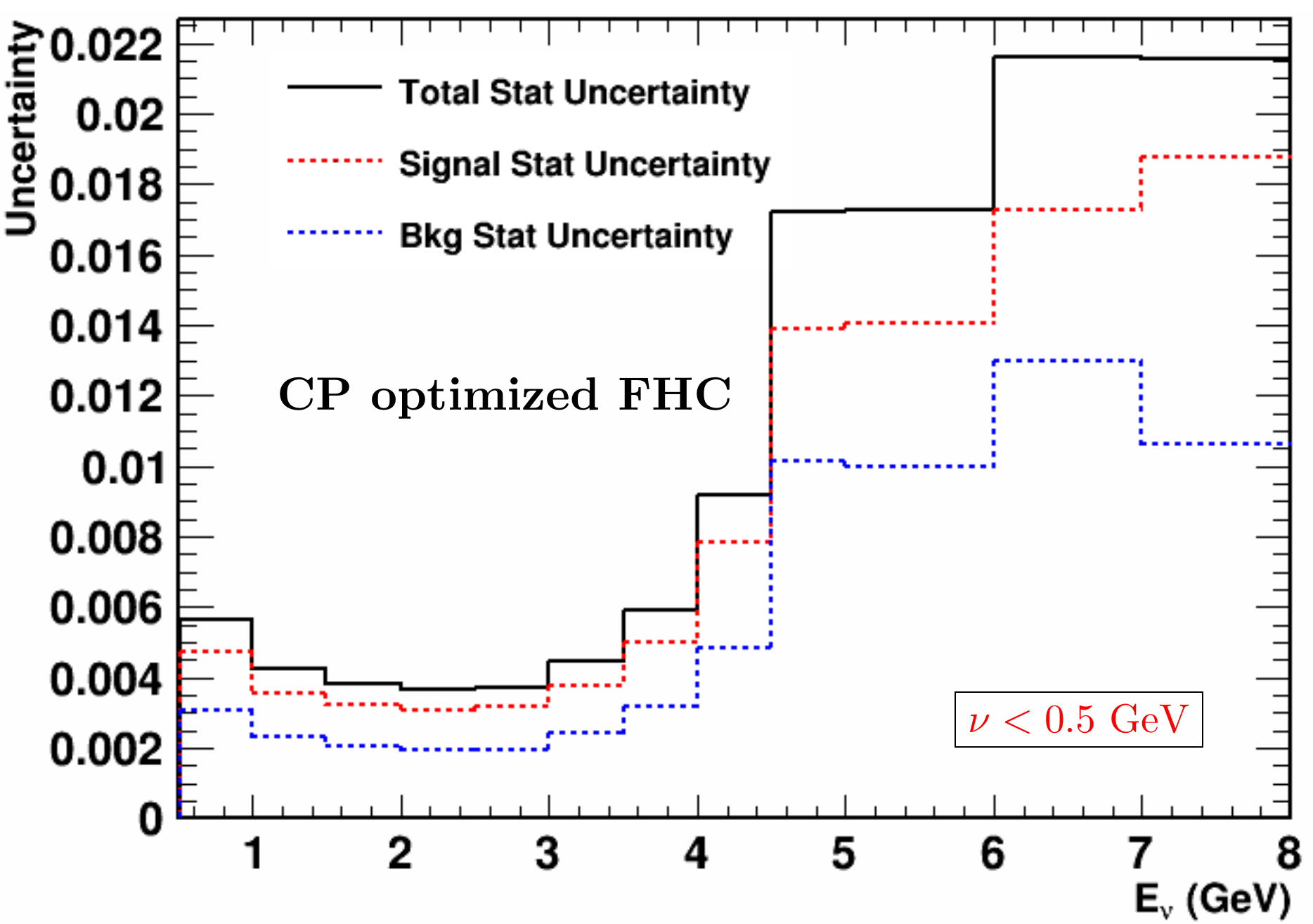}
\end{center}
\caption{Effect of the C background subtraction using the dedicated graphite target
on the statistical uncertainty of the selected $\nu_\mu p \to \mu^- p \pi^+$ on H 
with $\nu<0.5$ GeV. See text for details. 
}
\label{fig:Hsel-BkgSubErr}
\end{figure}

The C background subtraction introduces an increase of the statistical uncertainties 
of the selected H samples, as discussed in Ref.~\cite{Duyang:2018lpe}.  
We checked the impact of this subtraction on the flux determinations 
described in Sec.~\ref{sec:results} with a detailed study of the corresponding energy 
dependence. Figure~\ref{fig:Hsel-BkgSubErr} summarizes our results 
for the $\nu_\mu p \to \mu^- p \pi^+$ flux sample on H.  The cut $\nu<0.5$ GeV 
increases the purity of this sample to about 94\%, since at small 
energy transfers C background events are more subject to nuclear effects, 
making the kinematic analysis more efficient. 
With the analysis and exposures of Ref.~\cite{Duyang:2018lpe} (low energy beam) 
we expect about 39,000 C background events to be subtracted from the flux sample. 
As a result, we obtain a modest increase in the statistical uncertainty of the 
H sample of about 20\% (Fig.~\ref{fig:Hsel-BkgSubErr}) compared to the 
ones shown in Fig.~\ref{fig:Hsel-NumuFlux}. 
We note that this statistical penalty can be further reduced by analytically 
smoothing the measured distributions from the graphite target and/or by using a 
tighter kinematic selection. 

\section{Summary} 
\label{sec:sum} 

We proposed a novel method to achieve a precise determination of 
relative and absolute $\nu_\mu$ and $\bar \nu_\mu$ fluxes 
using exclusive $\nu_\mu p \to \mu^- p \pi^+$, $\bar \nu_\mu p \to \mu^+ p \pi^-$, 
and $\bar \nu_\mu p \to \mu^+ n$ processes on hydrogen with small energy transfer $\nu$. 
These event topologies can be efficiently selected with the 
simple and safe technique we previously proposed, based upon the subtraction 
between dedicated CH$_2$ plastic and graphite (pure C) targets, 
embedded within a low-density high-resolution detector providing a control of the configuration, 
chemical composition and mass of the targets similar to electron scattering experiments. 

We performed a detailed study of the relevant experimental and model uncertainties 
in the proposed method for the flux determination. To this end, we considered a  
realistic case study with (anti)neutrino beams similar to the ones planned at the 
Long-Baseline Neutrino Facility at Fermilab. Our results show that relative (anti)neutrino 
fluxes can be measured to an overall accuracy better than 1\% in the main energy ranges 
-- including both statistical and systematic uncertainties -- 
with the selected $\nu (\bar \nu)$-H exclusive topologies. 
We also presented techniques to constrain all relevant systematic uncertainties using 
data themselves to minimize model dependencies. 
The analysis appears to be 
statistics limited and can be easily generalized to arbitrary (anti)neutrino input spectra. 
This level of accuracy cannot be achieved by other techniques using nuclear targets.

\begin{acknowledgments}

We thank L. Alvarez-Ruso and G.T. Garvey for fruitful discussions. 
We thank M.V. Garzelli and C. Giunti for comments on the manuscript. 
This work was supported by Grant No. DE-SC0010073 from the Department of Energy, USA.

\end{acknowledgments}


\bibliography{main1}

\begin{thebibliography}{23}%
\makeatletter
\providecommand \@ifxundefined [1]{%
 \@ifx{#1\undefined}
}%
\providecommand \@ifnum [1]{%
 \ifnum #1\expandafter \@firstoftwo
 \else \expandafter \@secondoftwo
 \fi
}%
\providecommand \@ifx [1]{%
 \ifx #1\expandafter \@firstoftwo
 \else \expandafter \@secondoftwo
 \fi
}%
\providecommand \natexlab [1]{#1}%
\providecommand \enquote  [1]{``#1''}%
\providecommand \bibnamefont  [1]{#1}%
\providecommand \bibfnamefont [1]{#1}%
\providecommand \citenamefont [1]{#1}%
\providecommand \href@noop [0]{\@secondoftwo}%
\providecommand \href [0]{\begingroup \@sanitize@url \@href}%
\providecommand \@href[1]{\@@startlink{#1}\@@href}%
\providecommand \@@href[1]{\endgroup#1\@@endlink}%
\providecommand \@sanitize@url [0]{\catcode `\\12\catcode `\$12\catcode
  `\&12\catcode `\#12\catcode `\^12\catcode `\_12\catcode `\%12\relax}%
\providecommand \@@startlink[1]{}%
\providecommand \@@endlink[0]{}%
\providecommand \url  [0]{\begingroup\@sanitize@url \@url }%
\providecommand \@url [1]{\endgroup\@href {#1}{\urlprefix }}%
\providecommand \urlprefix  [0]{URL }%
\providecommand \Eprint [0]{\href }%
\providecommand \doibase [0]{http://dx.doi.org/}%
\providecommand \selectlanguage [0]{\@gobble}%
\providecommand \bibinfo  [0]{\@secondoftwo}%
\providecommand \bibfield  [0]{\@secondoftwo}%
\providecommand \translation [1]{[#1]}%
\providecommand \BibitemOpen [0]{}%
\providecommand \bibitemStop [0]{}%
\providecommand \bibitemNoStop [0]{.\EOS\space}%
\providecommand \EOS [0]{\spacefactor3000\relax}%
\providecommand \BibitemShut  [1]{\csname bibitem#1\endcsname}%
\let\auto@bib@innerbib\@empty
\bibitem [{\citenamefont {Duyang}\ \emph {et~al.}(2018)\citenamefont {Duyang},
  \citenamefont {Guo}, \citenamefont {Mishra},\ and\ \citenamefont
  {Petti}}]{Duyang:2018lpe}%
  \BibitemOpen
  \bibfield  {author} {\bibinfo {author} {\bibfnamefont {H.}~\bibnamefont
  {Duyang}}, \bibinfo {author} {\bibfnamefont {B.}~\bibnamefont {Guo}},
  \bibinfo {author} {\bibfnamefont {S.~R.}\ \bibnamefont {Mishra}}, \ and\
  \bibinfo {author} {\bibfnamefont {R.}~\bibnamefont {Petti}},\ }\href@noop {}
  {\  (\bibinfo {year} {2018})},\ \Eprint {http://arxiv.org/abs/1809.08752}
  {arXiv:1809.08752 [hep-ph]} \BibitemShut {NoStop}%
\bibitem [{\citenamefont {Juszczak}\ \emph {et~al.}(2006)\citenamefont
  {Juszczak}, \citenamefont {Nowak},\ and\ \citenamefont
  {Sobczyk}}]{Juszczak:2005zs}%
  \BibitemOpen
  \bibfield  {author} {\bibinfo {author} {\bibfnamefont {C.}~\bibnamefont
  {Juszczak}}, \bibinfo {author} {\bibfnamefont {J.~A.}\ \bibnamefont {Nowak}},
  \ and\ \bibinfo {author} {\bibfnamefont {J.~T.}\ \bibnamefont {Sobczyk}},\
  }\bibfield  {booktitle} {\emph {\bibinfo {booktitle} {{NuInt05, proceedings
  of the 4th International Workshop on Neutrino-Nucleus Interactions in the
  Few-GeV Region, Okayama, Japan, 26-29 September 2005}}},\ }\href {\doibase
  10.1016/j.nuclphysbps.2006.08.069} {\bibfield  {journal} {\bibinfo  {journal}
  {Nucl. Phys. Proc. Suppl.}\ }\textbf {\bibinfo {volume} {159}},\ \bibinfo
  {pages} {211} (\bibinfo {year} {2006})},\ \bibinfo {note} {[,211(2005)]},\
  \Eprint {http://arxiv.org/abs/hep-ph/0512365} {arXiv:hep-ph/0512365 [hep-ph]}
  \BibitemShut {NoStop}%
\bibitem [{\citenamefont {Buss}\ \emph {et~al.}(2012)\citenamefont {Buss},
  \citenamefont {Gaitanos}, \citenamefont {Gallmeister}, \citenamefont {van
  Hees}, \citenamefont {Kaskulov}, \citenamefont {Lalakulich}, \citenamefont
  {Larionov}, \citenamefont {Leitner}, \citenamefont {Weil},\ and\
  \citenamefont {Mosel}}]{Buss:2011mx}%
  \BibitemOpen
  \bibfield  {author} {\bibinfo {author} {\bibfnamefont {O.}~\bibnamefont
  {Buss}}, \bibinfo {author} {\bibfnamefont {T.}~\bibnamefont {Gaitanos}},
  \bibinfo {author} {\bibfnamefont {K.}~\bibnamefont {Gallmeister}}, \bibinfo
  {author} {\bibfnamefont {H.}~\bibnamefont {van Hees}}, \bibinfo {author}
  {\bibfnamefont {M.}~\bibnamefont {Kaskulov}}, \bibinfo {author}
  {\bibfnamefont {O.}~\bibnamefont {Lalakulich}}, \bibinfo {author}
  {\bibfnamefont {A.~B.}\ \bibnamefont {Larionov}}, \bibinfo {author}
  {\bibfnamefont {T.}~\bibnamefont {Leitner}}, \bibinfo {author} {\bibfnamefont
  {J.}~\bibnamefont {Weil}}, \ and\ \bibinfo {author} {\bibfnamefont
  {U.}~\bibnamefont {Mosel}},\ }\href {\doibase 10.1016/j.physrep.2011.12.001}
  {\bibfield  {journal} {\bibinfo  {journal} {Phys. Rept.}\ }\textbf {\bibinfo
  {volume} {512}},\ \bibinfo {pages} {1} (\bibinfo {year} {2012})},\ \Eprint
  {http://arxiv.org/abs/1106.1344} {arXiv:1106.1344 [hep-ph]} \BibitemShut
  {NoStop}%
\bibitem [{\citenamefont {Andreopoulos}\ \emph {et~al.}(2010)\citenamefont
  {Andreopoulos} \emph {et~al.}}]{Andreopoulos:2009rq}%
  \BibitemOpen
  \bibfield  {author} {\bibinfo {author} {\bibfnamefont {C.}~\bibnamefont
  {Andreopoulos}} \emph {et~al.},\ }\href {\doibase 10.1016/j.nima.2009.12.009}
  {\bibfield  {journal} {\bibinfo  {journal} {Nucl. Instrum. Meth.}\ }\textbf
  {\bibinfo {volume} {A614}},\ \bibinfo {pages} {87} (\bibinfo {year}
  {2010})},\ \Eprint {http://arxiv.org/abs/0905.2517} {arXiv:0905.2517
  [hep-ph]} \BibitemShut {NoStop}%
\bibitem [{\citenamefont {Acciarri}\ \emph {et~al.}(2015)\citenamefont
  {Acciarri} \emph {et~al.}}]{Acciarri:2015uup}%
  \BibitemOpen
  \bibfield  {author} {\bibinfo {author} {\bibfnamefont {R.}~\bibnamefont
  {Acciarri}} \emph {et~al.} (\bibinfo {collaboration} {DUNE}),\ }\href@noop {}
  {\  (\bibinfo {year} {2015})},\ \Eprint {http://arxiv.org/abs/1512.06148}
  {arXiv:1512.06148 [physics.ins-det]} \BibitemShut {NoStop}%
\bibitem [{\citenamefont {Acciarri}\ \emph {et~al.}(2016)\citenamefont
  {Acciarri} \emph {et~al.}}]{Acciarri:2016ooe}%
  \BibitemOpen
  \bibfield  {author} {\bibinfo {author} {\bibfnamefont {R.}~\bibnamefont
  {Acciarri}} \emph {et~al.} (\bibinfo {collaboration} {DUNE}),\ }\href@noop {}
  {\  (\bibinfo {year} {2016})},\ \Eprint {http://arxiv.org/abs/1601.02984}
  {arXiv:1601.02984 [physics.ins-det]} \BibitemShut {NoStop}%
\bibitem [{\citenamefont {Agostinelli}\ \emph {et~al.}(2003)\citenamefont
  {Agostinelli} \emph {et~al.}}]{Agostinelli:2002hh}%
  \BibitemOpen
  \bibfield  {author} {\bibinfo {author} {\bibfnamefont {S.}~\bibnamefont
  {Agostinelli}} \emph {et~al.} (\bibinfo {collaboration} {GEANT4}),\ }\href
  {\doibase 10.1016/S0168-9002(03)01368-8} {\bibfield  {journal} {\bibinfo
  {journal} {Nucl. Instrum. Meth.}\ }\textbf {\bibinfo {volume} {A506}},\
  \bibinfo {pages} {250} (\bibinfo {year} {2003})}\BibitemShut {NoStop}%
\bibitem [{\citenamefont {Altegoer}\ \emph {et~al.}(1998)\citenamefont
  {Altegoer} \emph {et~al.}}]{Altegoer:1997gv}%
  \BibitemOpen
  \bibfield  {author} {\bibinfo {author} {\bibfnamefont {J.}~\bibnamefont
  {Altegoer}} \emph {et~al.} (\bibinfo {collaboration} {NOMAD}),\ }\href
  {\doibase 10.1016/S0168-9002(97)01079-6} {\bibfield  {journal} {\bibinfo
  {journal} {Nucl. Instrum. Meth.}\ }\textbf {\bibinfo {volume} {A404}},\
  \bibinfo {pages} {96} (\bibinfo {year} {1998})}\BibitemShut {NoStop}%
\bibitem [{\citenamefont {Mishra}(1990)}]{Mishra90}%
  \BibitemOpen
  \bibfield  {author} {\bibinfo {author} {\bibfnamefont {S.~R.}\ \bibnamefont
  {Mishra}},\ }\href@noop {} {\bibfield  {journal} {\bibinfo  {journal}
  {Proceedings of the Workshop on Hadron Structure Functions and Parton
  Distributions (edited by D. Geesaman et al., World Scientific)}\ ,\ \bibinfo
  {pages} {84}} (\bibinfo {year} {1990})}\BibitemShut {NoStop}%
\bibitem [{\citenamefont {Bodek}\ \emph {et~al.}(2012)\citenamefont {Bodek},
  \citenamefont {Sarica}, \citenamefont {Naples},\ and\ \citenamefont
  {Ren}}]{Bodek:2012uu}%
  \BibitemOpen
  \bibfield  {author} {\bibinfo {author} {\bibfnamefont {A.}~\bibnamefont
  {Bodek}}, \bibinfo {author} {\bibfnamefont {U.}~\bibnamefont {Sarica}},
  \bibinfo {author} {\bibfnamefont {D.}~\bibnamefont {Naples}}, \ and\ \bibinfo
  {author} {\bibfnamefont {L.}~\bibnamefont {Ren}},\ }\href {\doibase
  10.1140/epjc/s10052-012-1973-6} {\bibfield  {journal} {\bibinfo  {journal}
  {Eur. Phys. J.}\ }\textbf {\bibinfo {volume} {C72}},\ \bibinfo {pages} {1973}
  (\bibinfo {year} {2012})},\ \Eprint {http://arxiv.org/abs/1201.3025}
  {arXiv:1201.3025 [hep-ex]} \BibitemShut {NoStop}%
\bibitem [{\citenamefont {Kulagin}\ and\ \citenamefont
  {Petti}(2006)}]{Kulagin:2004ie}%
  \BibitemOpen
  \bibfield  {author} {\bibinfo {author} {\bibfnamefont {S.~A.}\ \bibnamefont
  {Kulagin}}\ and\ \bibinfo {author} {\bibfnamefont {R.}~\bibnamefont
  {Petti}},\ }\href {\doibase 10.1016/j.nuclphysa.2005.10.011} {\bibfield
  {journal} {\bibinfo  {journal} {Nucl. Phys.}\ }\textbf {\bibinfo {volume}
  {A765}},\ \bibinfo {pages} {126} (\bibinfo {year} {2006})},\ \Eprint
  {http://arxiv.org/abs/hep-ph/0412425} {arXiv:hep-ph/0412425 [hep-ph]}
  \BibitemShut {NoStop}%
\bibitem [{\citenamefont {Kulagin}\ and\ \citenamefont
  {Petti}(2007)}]{Kulagin:2007ju}%
  \BibitemOpen
  \bibfield  {author} {\bibinfo {author} {\bibfnamefont {S.~A.}\ \bibnamefont
  {Kulagin}}\ and\ \bibinfo {author} {\bibfnamefont {R.}~\bibnamefont
  {Petti}},\ }\href {\doibase 10.1103/PhysRevD.76.094023} {\bibfield  {journal}
  {\bibinfo  {journal} {Phys. Rev.}\ }\textbf {\bibinfo {volume} {D76}},\
  \bibinfo {pages} {094023} (\bibinfo {year} {2007})},\ \Eprint
  {http://arxiv.org/abs/hep-ph/0703033} {arXiv:hep-ph/0703033 [HEP-PH]}
  \BibitemShut {NoStop}%
\bibitem [{\citenamefont {Kulagin}\ and\ \citenamefont
  {Petti}(2014)}]{Kulagin:2014vsa}%
  \BibitemOpen
  \bibfield  {author} {\bibinfo {author} {\bibfnamefont {S.~A.}\ \bibnamefont
  {Kulagin}}\ and\ \bibinfo {author} {\bibfnamefont {R.}~\bibnamefont
  {Petti}},\ }\href {\doibase 10.1103/PhysRevC.90.045204} {\bibfield  {journal}
  {\bibinfo  {journal} {Phys. Rev.}\ }\textbf {\bibinfo {volume} {C90}},\
  \bibinfo {pages} {045204} (\bibinfo {year} {2014})},\ \Eprint
  {http://arxiv.org/abs/1405.2529} {arXiv:1405.2529 [hep-ph]} \BibitemShut
  {NoStop}%
\bibitem [{\citenamefont {Alvarez-Ruso}\ \emph {et~al.}(2018)\citenamefont
  {Alvarez-Ruso} \emph {et~al.}}]{Alvarez-Ruso:2017oui}%
  \BibitemOpen
  \bibfield  {author} {\bibinfo {author} {\bibfnamefont {L.}~\bibnamefont
  {Alvarez-Ruso}} \emph {et~al.},\ }\href {\doibase 10.1016/j.ppnp.2018.01.006}
  {\bibfield  {journal} {\bibinfo  {journal} {Prog. Part. Nucl. Phys.}\
  }\textbf {\bibinfo {volume} {100}},\ \bibinfo {pages} {1} (\bibinfo {year}
  {2018})},\ \Eprint {http://arxiv.org/abs/1706.03621} {arXiv:1706.03621
  [hep-ph]} \BibitemShut {NoStop}%
\bibitem [{\citenamefont {Nakamura}\ \emph {et~al.}(2015)\citenamefont
  {Nakamura}, \citenamefont {Kamano},\ and\ \citenamefont
  {Sato}}]{Nakamura:2015rta}%
  \BibitemOpen
  \bibfield  {author} {\bibinfo {author} {\bibfnamefont {S.~X.}\ \bibnamefont
  {Nakamura}}, \bibinfo {author} {\bibfnamefont {H.}~\bibnamefont {Kamano}}, \
  and\ \bibinfo {author} {\bibfnamefont {T.}~\bibnamefont {Sato}},\ }\href
  {\doibase 10.1103/PhysRevD.92.074024} {\bibfield  {journal} {\bibinfo
  {journal} {Phys. Rev.}\ }\textbf {\bibinfo {volume} {D92}},\ \bibinfo {pages}
  {074024} (\bibinfo {year} {2015})},\ \Eprint
  {http://arxiv.org/abs/1506.03403} {arXiv:1506.03403 [hep-ph]} \BibitemShut
  {NoStop}%
\bibitem [{\citenamefont {Hernandez}\ and\ \citenamefont
  {Nieves}(2017)}]{Hernandez:2016yfb}%
  \BibitemOpen
  \bibfield  {author} {\bibinfo {author} {\bibfnamefont {E.}~\bibnamefont
  {Hernandez}}\ and\ \bibinfo {author} {\bibfnamefont {J.}~\bibnamefont
  {Nieves}},\ }\href@noop {} {\bibfield  {journal} {\bibinfo  {journal} {Phys.
  Rev.}\ }\textbf {\bibinfo {volume} {D95}},\ \bibinfo {pages} {053007}
  (\bibinfo {year} {2017})},\ \Eprint {http://arxiv.org/abs/1612.02343}
  {arXiv:1612.02343 [hep-ph]} \BibitemShut {NoStop}%
\bibitem [{\citenamefont {Bhattacharya}\ \emph {et~al.}(2011)\citenamefont
  {Bhattacharya}, \citenamefont {Hill},\ and\ \citenamefont
  {Paz}}]{Bhattacharya:2011ah}%
  \BibitemOpen
  \bibfield  {author} {\bibinfo {author} {\bibfnamefont {B.}~\bibnamefont
  {Bhattacharya}}, \bibinfo {author} {\bibfnamefont {R.~J.}\ \bibnamefont
  {Hill}}, \ and\ \bibinfo {author} {\bibfnamefont {G.}~\bibnamefont {Paz}},\
  }\href {\doibase 10.1103/PhysRevD.84.073006} {\bibfield  {journal} {\bibinfo
  {journal} {Phys. Rev.}\ }\textbf {\bibinfo {volume} {D84}},\ \bibinfo {pages}
  {073006} (\bibinfo {year} {2011})},\ \Eprint {http://arxiv.org/abs/1108.0423}
  {arXiv:1108.0423 [hep-ph]} \BibitemShut {NoStop}%
\bibitem [{\citenamefont {Park}\ \emph {et~al.}(2016)\citenamefont {Park} \emph
  {et~al.}}]{Park:2015eqa}%
  \BibitemOpen
  \bibfield  {author} {\bibinfo {author} {\bibfnamefont {J.}~\bibnamefont
  {Park}} \emph {et~al.} (\bibinfo {collaboration} {MINERvA}),\ }\href
  {\doibase 10.1103/PhysRevD.93.112007} {\bibfield  {journal} {\bibinfo
  {journal} {Phys. Rev.}\ }\textbf {\bibinfo {volume} {D93}},\ \bibinfo {pages}
  {112007} (\bibinfo {year} {2016})},\ \Eprint
  {http://arxiv.org/abs/1512.07699} {arXiv:1512.07699 [physics.ins-det]}
  \BibitemShut {NoStop}%
\bibitem [{\citenamefont {Yao}\ \emph {et~al.}(2018)\citenamefont {Yao},
  \citenamefont {Alvarez-Ruso}, \citenamefont {Hiller~Blin},\ and\
  \citenamefont {Vicente~Vacas}}]{Yao:2018pzc}%
  \BibitemOpen
  \bibfield  {author} {\bibinfo {author} {\bibfnamefont {D.-L.}\ \bibnamefont
  {Yao}}, \bibinfo {author} {\bibfnamefont {L.}~\bibnamefont {Alvarez-Ruso}},
  \bibinfo {author} {\bibfnamefont {A.~N.}\ \bibnamefont {Hiller~Blin}}, \ and\
  \bibinfo {author} {\bibfnamefont {M.~J.}\ \bibnamefont {Vicente~Vacas}},\
  }\href {\doibase 10.1103/PhysRevD.98.076004} {\bibfield  {journal} {\bibinfo
  {journal} {Phys. Rev.}\ }\textbf {\bibinfo {volume} {D98}},\ \bibinfo {pages}
  {076004} (\bibinfo {year} {2018})},\ \Eprint
  {http://arxiv.org/abs/1806.09364} {arXiv:1806.09364 [hep-ph]} \BibitemShut
  {NoStop}%
\bibitem [{\citenamefont {Alvarez-Ruso}()}]{Luis19}%
  \BibitemOpen
  \bibfield  {author} {\bibinfo {author} {\bibfnamefont {L.}~\bibnamefont
  {Alvarez-Ruso}},\ }\href@noop {} {}\bibinfo {howpublished} {Private
  communication}\BibitemShut {NoStop}%
\bibitem [{\citenamefont {Petti}()}]{Petti2018}%
  \BibitemOpen
  \bibfield  {author} {\bibinfo {author} {\bibfnamefont {R.}~\bibnamefont
  {Petti}},\ }\href
  {https://indico.cern.ch/event/721473/contributions/3034869/} {\ }\bibinfo
  {note} {Workshop on Near Detector Physics at Neutrino Experiments, CERN,
  18-22 June 2018,
  \url{https://indico.cern.ch/event/721473/contributions/3034869/}}\BibitemShut
  {NoStop}%
\bibitem [{\citenamefont {Bernardini}\ \emph {et~al.}()\citenamefont
  {Bernardini} \emph {et~al.}}]{ESGprop}%
  \BibitemOpen
  \bibfield  {author} {\bibinfo {author} {\bibfnamefont {P.}~\bibnamefont
  {Bernardini}} \emph {et~al.},\ }\href
  {https://indico.cern.ch/event/765096/contributions/3295805/} {\ }\bibinfo
  {note} {Proposal accepted for the European Particle Physics Strategy Update
  2018-2020,
  \url{https://indico.cern.ch/event/765096/contributions/3295805/}}\BibitemShut
  {NoStop}%
\bibitem [{\citenamefont {Wu}\ \emph {et~al.}(2008)\citenamefont {Wu} \emph
  {et~al.}}]{Wu:2007ab}%
  \BibitemOpen
  \bibfield  {author} {\bibinfo {author} {\bibfnamefont {Q.}~\bibnamefont {Wu}}
  \emph {et~al.} (\bibinfo {collaboration} {NOMAD}),\ }\href {\doibase
  10.1016/j.physletb.2007.12.027} {\bibfield  {journal} {\bibinfo  {journal}
  {Phys. Lett.}\ }\textbf {\bibinfo {volume} {B660}},\ \bibinfo {pages} {19}
  (\bibinfo {year} {2008})},\ \Eprint {http://arxiv.org/abs/0711.1183}
  {arXiv:0711.1183 [hep-ex]} \BibitemShut {NoStop}%
\end{thebibliography}%

\end{document}